\documentclass{llncs}
\usepackage{llncsdoc}
\usepackage{epsfig}
\usepackage{epstopdf}
\usepackage{hyperref}
\usepackage{cite}
\usepackage{array}
\usepackage{multirow}
\usepackage[table]{xcolor}

\title{Effects of Coupling in\\Human-virtual Agent Body Interaction}
\author{Elisabetta Bevacqua, Igor Stankovi\'{c}, Ayoub Maatallaoui, \\Alexis N\'{e}d\'{e}lec, \and Pierre De Loor}
\institute{UEB, Lab-STICC, ENIB\\
\email{\{bevacqua,stankovic,maatallaoui,nedelec,deloor\}@enib.fr}
}

\begin{document}
\maketitle
\begin{abstract} 
This paper presents a study of the dynamic coupling between a user and a virtual character during body interaction. Coupling is directly linked with other dimensions, such as co-presence, engagement, and believability, and was measured in an experiment that allowed users to describe their subjective feelings about those dimensions of interest. The experiment was based on a theatrical game involving the imitation of slow upper-body movements and the proposal of new movements by the user and virtual agent. The agent's behaviour varied in autonomy: the agent could limit itself to imitating the user's movements only, initiate new movements, or combine both behaviours. After the game, each participant completed a questionnaire regarding their engagement in the interaction, their subjective feeling about the co-presence of the agent, etc. Based on four main dimensions of interest, we tested several hypotheses against our experimental results, which are discussed here.

\end{abstract}

\section{Introduction}
\label{introduction}
\parskip 0pt
\textit{Coupling} \cite{deloor2014} is the continuous mutual influence between two individuals, and has a dynamic specific to the dyad. It possesses the capability to resist disturbance, and compensates by evolving the interaction. Disturbances come from both the environment and from within the individuals, depending on how they perceive the interaction. This definition is recursive since coupling exists because of the human effort to ``recover'' it as its quality decreases; this is why it is highly complex to reproduce when employing virtual agents. Coupling between two persons implies an \textit{evolving equilibrium} between regularity and surprise, and it is a fundamental key to establish an interaction. Our assumption is that coupling and sense-making are tightly linked to a subjective feeling of several dimensions of interaction. In this paper we focus on co-presence, believability, and engagement as these are important dimensions frequently addressed by the virtual character community. 

Many studies in the field of human-agent interaction have tried to develop believable, co-present, and/or engaging agents. If presence is addressed in virtual reality as the feeling of ``being there'' \cite{heeter1992}, co-presence is the feeling of ``being with'' \cite{bailenson2004}. Believability is how an object or character fits a user's model, and engagement is a measure for being ``into the game''. The improvement of these subjective feelings must address two problems. The first is the multi-dimensionality of the interaction. Emotional feedback, expressed through facial expressions, is just one of the cues that helps agents build a better rapport with humans \cite{wong2012}. Also, back-channels are considered ``the most accessible example of the real-time responsiveness that underpins many successful interpersonal interactions'', and expressive feedback, such as a nod or an ``a-ha'' (which literally means ``I am listening, tell me more''), given at the right moment, heightens the degree of convergence \cite{traum2012}. In addition, synchrony is also an important parameter in human-agent coupling \cite{prepin2013}. The second problem is the difficulty of defining and evaluating the subjective feelings of users. There is much debate on the link between feeling, in the sense of ``What is it like?'', and physiological responses \cite{insko2003}. The debate about the notion of presence is well known \cite{schuemie2001}. Some researchers argue that co-presence is primarily subjective, so they try to define a ``good'' subjective questionnaire \cite{witmer1998}, while others stress that only physiological measures can provide progress on the understanding of presence \cite{slater2004}. It is also possible to find objective measures for believability \cite{Riedl2005}, or to use subjective evaluation techniques \cite{Hingston2012}, while engagement can be evaluated by feeling (e.g. of pleasure, or control), or through objective measures in terms of time before fatigue \cite{obrien2008}.

To study the links between coupling and the three dimensions (believability, co-presence, and engagement), we propose a body interaction experiment that allows us to vary the coupling between a human and a virtual character. We aim at improving the interaction experience by gathering insights into the principles necessary for implementing virtual characters. Additionally, the experiment could help us to better understand how ``subjective feeling'' should be evaluated.

Details on the experiment, its variations (the different condition scenarios), and tested hypotheses are given in Sect. \ref{experiment}, while Sect. \ref{method} explains the methods utilized. Section \ref{result} presents several result sets, which are discussed in Sect. \ref{discussion}. Conclusions are drawn in the final section (Sect. \ref{conclusion}).

\section{Experiment}
\label{experiment}
An evaluation test was used to assess the dynamic coupling between a human user and virtual agent. In a theatrical exercise, two players facing each other imitated the other person's upper-body movements but introduced subtle changes by proposing, from time to time, new movements. This dyadic imitation game causes dynamic notions of coupling and interaction to emerge naturally from both players. Regularity (through the imitation of the other subject) and surprise (seen in the new movements) are intrinsic to the exercise, and are perfectly balanced. 
This game meets our needs perfectly, and the participants were asked to play it with a virtual agent. 

Our system uses a motion capture device (Microsoft Kinect) to collect positional information about the user's body. Captured coordinates are passed through a simple averaging filter to reduce noise, and then sent onto a synthesis module built in Unity3D. It uses the body coordinates and inverse kinematics to make the virtual agent strictly imitate/follow the user's movements. A Wizard of Oz (WOZ) technique allows an agent to create new movements during the interaction. The WOZ, managed by one of the evaluators through keyboard controls, can change the agent's hands directions. No blending issues between old and new movements were perceivable since changes were quite slow. When the WOZ is disabled, the agent will again start strictly following the user's movements. To study only body movements and for artistic reasons, one of Joan Mir\'{o}'s colourful paintings involving a devil-like minimalist character made of black segments, was utilized as the agent (see Fig. \ref{fig1}). Participants interacted with the agent by utilizing one of three scenario conditions:

\begin{figure}
\centering
\includegraphics[scale=0.24]{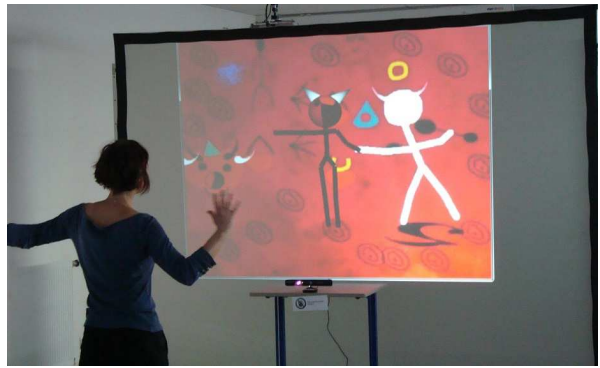}
\caption{Scene installation: When the user is detected, the devil-like character ``jumps out'' of the painting leaving a white empty shape, and the interaction starts}
\label{fig1}
\end{figure}

\begin{itemize}
\renewcommand{\labelitemi}{$\bullet$}
   \item \emph{$1^{st}$ condition (C1):} the agent's behaviour was a pure imitation of the user's.
   \item \emph{$2^{nd}$ condition (C2):} the agent's behaviour was partially driven by the WOZ.
   \item \emph{$3^{rd}$ condition (C3):} the agent's movements were controlled by a previously recorded motion capture file of another person playing the game.
\end{itemize}

The agent imitates the user's movement in cases C1 and C2 with a slight delay. Without such a delay, the agent almost instantly imitates the user's behaviour, which does not seem natural to the human participant, and makes the agent seem too obviously computer-driven. Pretests showed that employing a half second delay is a good solution to this problem.

Prior to our experiment, we formulated several hypotheses involving four dimensions of interest (coupling, co-presence, engagement, and believability) that will be measured through a questionnaire:

\begin{description}
\item[Hypothesis 1.] 
The four dimensions would be most prominent in condition C2 rather than in C1 or C3. Also, since the agent does not react to human behaviour in C3, 
no connection would develop between the subject and the agent. This suggests that higher results would be expected in C1 than in C3.

\item[Hypothesis 2.] Level of engagement, sense of co-presence, and believability are due to a subtle equilibrium between surprise and regularity during an interaction. In other words, co-presence, engagement, and believability are connected to the level of coupling.

\item[Hypothesis 3.] Engagement and a feeling of co-presence are linked. Perceiving the co-presence of the agent makes the game more fun, and so more engaging.
\end{description}

\section{Method}
\label{method}
The experiment was conducted at a school during an exhibition about the links between art and science. We decided on an independent-measures design: each subject participated in just one condition scenario (C1, C2, or C3). Data from forty-one French-speaking subjects (20\% women, 80\% men) was collected: thirteen subjects, age from 18 to 30 ($Median=21$), participated in C1; fifteen, age from 15 to 42 ($Median=21$), participated in C2; thirteen participants, age from 19 to 46 ($Median=20$), interacted with the agent under condition C3. 

The exercise was explained to the subjects and they were invited to play the game with one of the evaluators. This introduction encouraged the participants to feel the type of connections that could occur in the game. The subjects did not know which condition they were playing, and to measure their level of engagement, no time limit was imposed. At the end of the interaction, each participant filled in a questionnaire  (see Table \ref{tab-questions}) to judge their experience and the agent's behaviour.

\begin{table}
\caption{The sixteen statements in our questionnaire}
\begin{center}
  \begin{tabular}{|l|l|}
\hline	
\textbf{Dimension} & \textbf{Question} \\ \hline\hline
\emph{Coupling} & \textbf{q1.} I had the impression that the agent was proposing new movements. \\ 
		  & \textbf{q2.} I had the impression that the agent was following my movements. \\ 
		  & \textbf{q3.} I had the feeling that the agent's behaviour was connected to mine. \\ 
		  & \textbf{q4.} The agent did not take my movements into account. \\ 
		  & \textbf{q5.} I was able to make the agent follow me. \\ 
	  	  & \textbf{q6.} I was surprised by the agent's behaviour. \\ \hline
\emph{Co-presence} & \textbf{q7.} I had the impression that I was in the presence of another being. \\ 
		  & \textbf{q8.} I had the feeling that the agent was aware of my presence. \\ 
		  & \textbf{q9.} I perceived the agent as a simple computer program. \\ 
		  & \textbf{q10.} The agent seemed aware of its own behaviour. \\ \hline
\emph{Engagement} & \textbf{q11.} I enjoyed playing with the agent. \\ 
		    & \textbf{q12.}  I had the feeling that I was really playing with the agent. \\ 
		    & \textbf{q13.} Playing the game with the agent was easy. \\ \hline
\emph{Believability} 	& \textbf{q14.} The agent's behaviour made me think of human behaviour. \\ 
	 		& \textbf{q15.} I don't think that the agent was behaving like a real person. \\ 
	 		& \textbf{q16.} I had the impression that the agent was controlled by a human. \\ \hline	
  \end{tabular}
\end{center}
\label{tab-questions}
\end{table}

The questionnaire contained sixteen statements (each used a 6-point Likert scale: 1 = disagree strongly; 6 = agree strongly) grouped according to the four dimensions we have retained. Six of them are based on the definition of coupling presented in \cite{deloor2014} and then they are related to the feeling of regularities and surprises during the interaction. Our evaluation of co-presence drew inspiration from a questionnaire proposed in \cite{bailenson2004}. Level of engagement was evaluated according to how enjoyable the agent interaction was for the participant, the ease of the interaction, and whether the user felt involved in the game. We also recorded the length of each interaction with the aim of collecting additional information on the users' engagement since more engaging interactions last longer. To assess the perceived believability of the agent's behaviour, the questionnaire addressed the closeness of the agent's behaviour to human actions.

\section{Results} 
\label{result}
Each questionnaire was analysed by evaluating each statement within the context of the three condition scenarios (C1, C2, and C3). We compared the answers to each question pairwise, by considering each pair of different conditions. For this we utilized the Wilcoxon test, a non-parametric equivalent of the \emph{t-test}. All our hypothesis analyses were one-tailed because the direction of each expected difference was specified. 
The results were significant for several of the statements, particularly for those that evaluated the feeling of coupling. Subjects easily recognized that the agent suggested fewer new movements (\emph{q1}) in condition C1 than in C2 ($p<.01$) or C3 ($p<.01$), and less in C2 than in C3 ($p<.01$). They also noticed when the agent was following the user more closely (\emph{q2}) in condition C1 rather than in C3 ($p<.01$), and more in C2 than in C3 ($p<.01$). Participants felt a stronger connection between their behaviour and the agent's (\emph{q3}) in C1 than in C3 ($p<.01$), and in C2 rather than in C3 ($p<.01$). In \emph{q4} (question 4), the agent was judged as taking the subject's behaviour more into account in condition C1 than in C3 ($p<.01$), and more in C2 more than in C3 ($p<.01$). The subjects were more surprised by the agent's behaviour (\emph{q6}) in C3 than in C1 ($p<.05$). 

\begin{figure}[h]
\centering
\includegraphics[width=10cm]{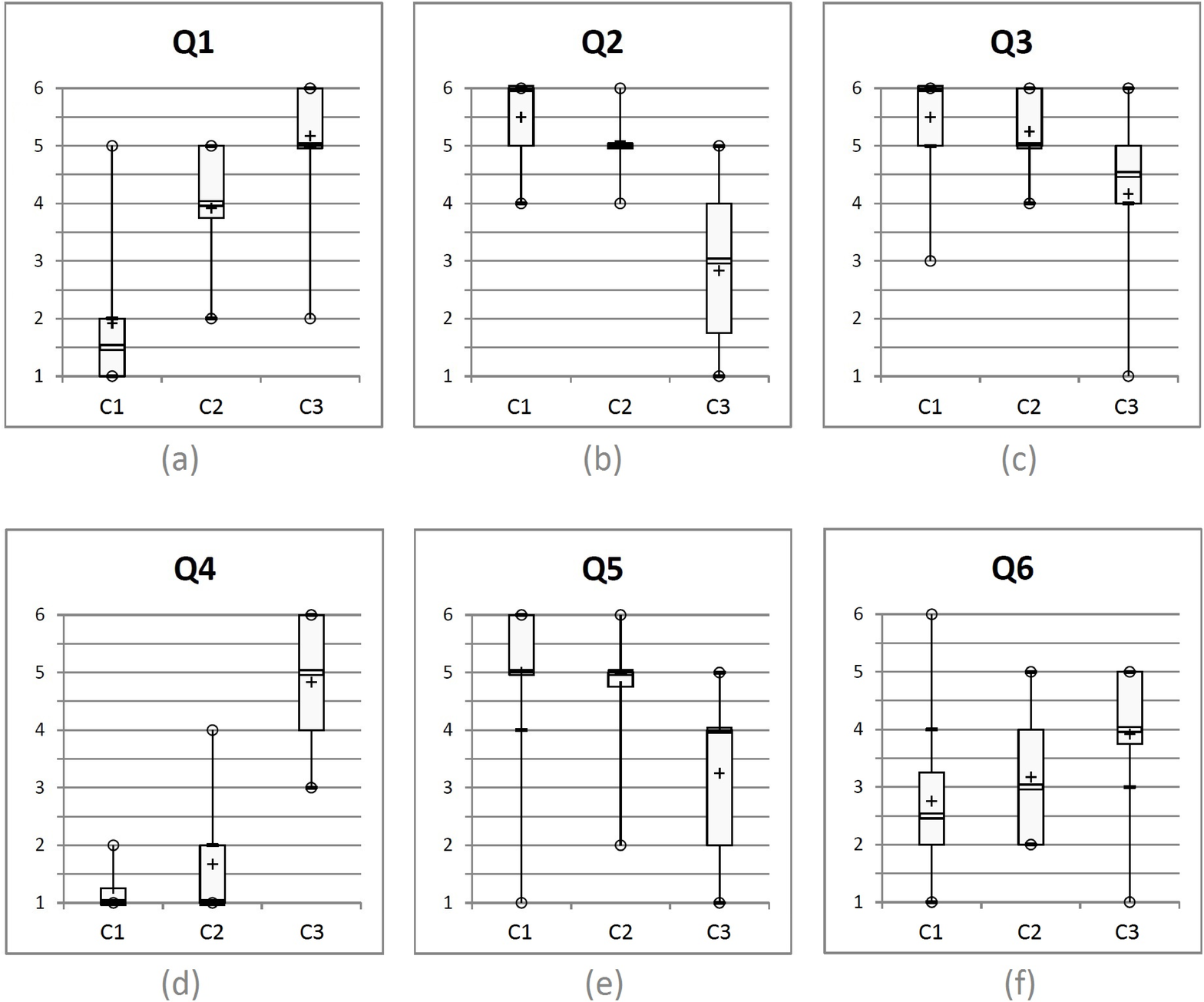}
\caption{Box plot diagrams of the coupling questions}\label{fig:plots}
\end{figure}

These results show that we did not find many significant differences between conditions C1 and C2. This is not surprising, particularly for those questions that asked the subjects if they felt that the agent was following them (\emph{q2} and \emph{q5}), or if they felt a connection with the agent (\emph{q3}), or if the agent was taking their movements into account (\emph{q4}), since the agent imitates the subjects in both conditions. However, the agent imitates less in condition C2, and people do tend to feel it, as shown in the box plots diagrams in Fig. \ref{fig:plots}. The diagrams of \emph{q2} and \emph{q5} shows that the subjects were more aware of the agent imitation in condition C1 than in C2. The diagram for \emph{q4} shows that people tend to believe that the agent takes their movements into account less in condition C2 than in C1. The agent seems a little more surprising in C2 than in C1, as shown by the box plot diagram of \emph{q6}. The diagram for \emph{q3} indicates that people feel slightly less connected to the agent in condition C2 than in C1. It is more surprising that the subjects also feel quite connected to the agent in condition C3 (even though there is a significant difference between the other two conditions). Perhaps this condition scenario forces people to try harder to play (since the agent doesn't interact at all), and the increased effort causes the players to imagine a connection that isn't there.

None of the questions regarding agent believability produced significant results, and the questions regarding co-presence contained almost no significant results. Participants felt that the agent was aware of its own behaviour (\emph{q10}) more in condition C3 than in C1 ($p<.01$) and C2 ($p<.05$), and more in C2 than in C1 ($p<.05$). No significant results were obtained for the feeling of engagement, except in question 12, where the subjects had the impression of playing with the agent more in condition C3 than in C1 ($p<.01$). 
 
Co-presence and engagement are very hard to evaluate solely through a questionnaire, although we did hope to find that a feeling of co-presence and engagement are linked to the level of coupling between the human and the virtual agent. Consequently, we built a correlation matrix between all the questions (disregarding the condition) by utilizing Spearman's Rho correlation coefficient. The two-tailed significance level of the correlation was measured to determine if it was significantly different from a zero-correlation in the positive or negative directions. Indeed, some of the questions were correlated (see Table \ref{tab-corr}). We also checked for a correlation between the questions and interaction duration, but only a weak result (Rho=0.268, $p<.05$) was obtained for question 12. It seems that the subjects interacted longer when they had a stronger impression of ``playing'' with the agent. Even without significant results for the agent's believability, correlations were detected between the questions on co-presence and believability and the questions on engagement and believability.

\begin{table}[h]
\caption{Results of Spearman's Rho correlation coefficient for N=41}
\begin{center}
  \begin{tabular}{|c|c|c|c||c|c|c|}
\hline	
& \textbf{Pair of}           & \textbf{Spearman's}  &             & \textbf{Pair of}      & \textbf{Spearman's}      &          \\ 
& \textbf{questions}         & \textbf{Rho}         &             & \textbf{questions}    & \textbf{Rho}            &          \\ \hline


\textbf{coupling-}  & q1-q10 & 0.757  & $p<.01$ & q4-q10 & 0.319  & $p<.05$  \\
\textbf{co-presence}  & q1-q8  & -0.296 & $p<.05$ & q5-q8  & 0.352  & $p<.05$  \\
	   & q2-q8  & 0.428  & $p<.01$ & q5-q10 & -0.299 & $p<.05$  \\ 
	   & q2-q10 & -0.407 & $p<.01$ & q6-q7  & 0.507  & $p<.01$  \\
	   & q3-q8  & 0.334  & $p<.05$ & q6-q10 & 0.306  & $p<.05$  \\
	   & q4-q8  & -0.324 & $p<.05$ &        &        &          \\ \hline

\textbf{coupling-}  & q1-q12 & 0.35   & $p<.05$ & q6-q11 & 0.311  & $p<.05$  \\ 
\textbf{engagement} & q3-q13 & 0.415  & $p<.01$ & q6-q12 & 0.383  & $p<.01$  \\ \hline 


\textbf{co-presence-}  & q7-q11 & 0.438 & $p<.01$ & q9-q11 & -0.408 & $p<.01$ \\ 
\textbf{engagement} & q7-q12 & 0.619 & $p<.01$ & q9-q12 & -0.29 & $p<.05$  \\ 
	   & q8-q11 & 0.283 & $p<.05$ & q10-q12 & 0.312 & $p<.05$ \\ \hline

\textbf{co-presence-}     & q7-q14  & 0.33  & $p<.05$ & q7-q15 & -0.375 & $p<.01$ \\ 
\textbf{believability} & q10-q14 & 0.481 & $p<.01$ &        &        &         \\ \hline

\textbf{engagement-}   & q11-q15 & -0.263   & $p<.05$ & q13-q14 & 0.475 & $p<.01$ \\ 
\textbf{believability} & q12-q14 & 0.323  & $p<.05$ & q13-q16 & 0.367 & $p<.05$ \\ 
	      & q15-q12 & -0.475 & $p<.01$ &         &       &          \\ \hline

\end{tabular}
\end{center}
\label{tab-corr}
\end{table}

\section{Discussion}
\label{discussion}
As for the first hypothesis, the experimental setup clearly allows coupling to emerge, and C2 gives the users the best feeling of coupling. Another property of C2 is its balance between surprise and regularity. For instance, \emph{q6} shows that C2 encourages more surprise than C1 and less than C3 (where the agent's behaviour is unconnected to the human's). Similarly, \emph{q1} shows that C2's behaviour is felt to lie somewhere between that of a passive agent (C1) and a directed agent (C3). A consideration of the questions concerning believability, co-presence, and engagement shows that C2 represents a balance between low and high autonomous behaviours (\emph{q10}). No other significant results concerning the discriminatory role of C2 were found, for which there are two possible explanations: 1.) feelings like co-presence and engagement are difficult to assess solely with a questionnaire, and 2.) some questions are victims of alternative interpretations, or were inadequate for discriminating our types of interaction. The second case was particularly true for the statements involving believability. All of our condition scenarios present agent behaviour intrinsically similar to that of a real human: the agent is solely driven by the user in C1, the agent partly reproduces the behaviour of the user in C2, and the agent plays back a behaviour generated by another human in condition C3. When considered this way, users can judge all the agent's behaviours to be human-like.

Condition C3 produced an interesting result. We did not expect any connection to be established between the subject and the agent, so initially thought that C1 and C2 would generate stronger engagements. However, \emph{q12}'s results show that users believe they are playing with the agent more in C3 than in C1. It appears that when a subject is imitating the agent, they also believe that they are playing the game together, and so feel an increased connection. The subjects seem to actively \emph{look for} this connection since it is the goal of the exercise. This indicates how a goal's role in this type of study can have a strong impact on the users' sense of engagement. There is a real difficulty in finding a balance between the fact that the subject must do something with the virtual character to induce coupling and how the user can become so focused on their role that the precise behaviour of the agent becomes less important. To test our last two hypotheses, we looked at the correlation between questions (independent of the experimental conditions). We examined the subjective links between human feelings, which are not necessarily related to the objective behaviour of the virtual character. Most of the questions about coupling correlate with one or more questions about co-presence (Table \ref{tab-corr}). From the user's point of view, the agent seems to be aware of the subjects' presence when it takes their movements into account and follows them, and this regularity is what people expect. Subjects feel the agent's presence strongly when its behaviour surprises them, so a balance between regularity and surprise increases a sense of co-presence. Even if people cannot objectively define what condition makes the feeling of co-presence stronger, they can subjectively feel that such a feeling has increased when coupling emerges. This confirms part of our second hypothesis. The other part, concerning the link between coupling and engagement, is harder to sustain since the only relevant correlations are between \emph{q11} and \emph{q6}, and between \emph{q12} and \emph{q6}. Clearly, surprise triggered by an agent's behaviour has an effect on a user's engagement because it increases the game's enjoyment, and heightens the impression of playing with the agent. This result may not be enough on its own to show a link between coupling and engagement, but it connects engagement with surprise as a component of coupling.

Table \ref{tab-corr} shows several correlations between questions about co-presence and those on engagement. When a subject enjoys playing with the agent, they feel more involved in the game, feel the agent's presence, and form an impression that the agent is perceiving them. As a consequence, the agent is not seen as a simple computer program. Although there is a clear link between the feeling of co-presence and engagement, the subjects do not find it easy to judge which condition scenario provokes a stronger sense of co-presence and engagement, but they do subjectively connect these two dimensions. Our hypotheses did not consider how the feelings of co-presence and engagement can influence the believability of agent behaviour, but a strong correlation between questions on co-presence and engagement, and questions on co-presence and believability was found (see Table \ref{tab-corr}). When users feel the agent's presence, or enjoy playing with it, they also judge its behaviour to be more human-like.

These correlations stress the link between subjective feelings and objective conditions. For example, C2 and C3 are objectively different, but no statistical difference was found between them regarding the feeling of co-presence. However, there are correlations between the feeling of co-presence and the feeling of coupling. This can be explained by how a person will construct a feeling of coupling with an agent even when such a connection does not really exist (see Fig. \ref{fig:plots}.c). For instance, the agent in C3 is constantly proposing new movements, but because the participants knows that the goal of the game is to imitate and be imitated, they try to create a (fake) coupling. They feel coupling because they want to, and once they think they are coupled with the agent, they also begin to feel its presence. To confirm this notion, we checked the correlations between coupling and co-presence in C3 and several interesting results were revealed. For example, \emph{q8} (``I had the feeling that the agent was aware of my presence'') correlates with almost all the questions on coupling: \emph{q1} (Rho=-0.541, $p<.05$), \emph{q2} (Rho=0.765, $p<.01$), \emph{q4} (Rho=-0.74, $p<.01$), and \emph{q5} (Rho=0.815, $p<.01$), for N=13; \emph{q9} and \emph{q4} also correlate (Rho=0.618, $p<.05$), indicating that the agent is perceived as less like a simple computer program when it takes the user's movements into account. The correlation between co-presence and coupling, which was found independent of the game condition, indicates that there is a ``hidden'' correlation phenomena at work. Perhaps there are two types of user: those who try to ``play the game'' by introducing a coupling, and so feel coupling and co-presence by the end of the experiment, and those users who do not ``play the game'' and so are denied those feelings.

\section{Conclusions}
\label{conclusion}
We have presented a study on human-virtual agent body interaction with four dimensions investigated: coupling, co-presence, engagement, and believability. Our results show that coupling is easily recognized by the participants, but the other three dimensions are harder to assess solely through a questionnaire. However, result correlations were found between coupling and co-presence, and between coupling and engagement. It seems that people do feel a sense of co-presence and heightened involvement when they feel coupled with the agent. There is a link between co-presence and coupling for subjects who make an effort to create coupling. This desire for interaction provokes a fake feeling of coupling which improves the feeling of co-presence. There may be an objective measure of this ``coupling willingness'', developing such a measure presents our next challenge.

Our results suggest that cognitive architectures must include coupling capabilities, as in \cite{prepin2013}, in agents intended to engender a strong sense of co-presence and engagement with users. As our work shows, when these two dimensional values are increased, then so does the believability of the agent behaviour. We also introduced the importance of interaction willingness and we propose a link between action and co-presence.

\subsubsection{Acknowledgement.} 
This work was funded by the ANR INGREDIBLE project: ANR-12-CORD-001 
(\url{http://www.ingredible.fr}).

\bibliographystyle{splncs}
\bibliography{iva14}

\end{document}